\documentclass[reprint,english,aip,apl,bibnotes]{revtex4-1}
\usepackage[latin9]{inputenc}
\usepackage{amstext}
\usepackage{graphicx}
\usepackage{amssymb}
\usepackage{multirow}
\usepackage{dcolumn}
\usepackage{color}
\usepackage{amsmath}
\usepackage{babel}
\usepackage{epstopdf}
\usepackage{natbib}
\usepackage{hyperref}
\usepackage{sidecap}
\definecolor{BLUE}{rgb}{0,0,1}
\hypersetup{
     colorlinks   = true,
     citecolor    = BLUE,
     urlcolor = BLUE,
     linkcolor = black
}

\makeatletter
\begin{document}
\title{Characterization and Reduction of Capacitive Loss Induced by Sub-Micron Josephson Junction Fabrication in Superconducting Qubits}

\author{A. Dunsworth}
\thanks{These authors contributed equally to this work}
\affiliation{Department of Physics, University of California, Santa Barbara, CA 93106-9530}
\author{A. Megrant}
\thanks{These authors contributed equally to this work}
\affiliation{Google, Santa Barbara, CA 93117}
\author{C. Quintana}
\affiliation{Department of Physics, University of California, Santa Barbara, CA 93106-9530}
\author{Zijun Chen}
\affiliation{Department of Physics, University of California, Santa Barbara, CA 93106-9530}
\author{R. Barends}
\affiliation{Google, Santa Barbara, CA 93117}
\author{B. Burkett}
\affiliation{Google, Santa Barbara, CA 93117}
\author{B. Foxen}
\affiliation{Department of Physics, University of California, Santa Barbara, CA 93106-9530}
\author{Yu Chen}
\affiliation{Google, Santa Barbara, CA 93117}
\author{B. Chiaro}
\affiliation{Department of Physics, University of California, Santa Barbara, CA 93106-9530}
\author{A. Fowler}
\affiliation{Google, Santa Barbara, CA 93117}
\author{R. Graff}
\affiliation{Google, Santa Barbara, CA 93117}
\author{E. Jeffrey}
\affiliation{Google, Santa Barbara, CA 93117}
\author{J. Kelly}
\affiliation{Google, Santa Barbara, CA 93117}
\author{E. Lucero}
\affiliation{Google, Santa Barbara, CA 93117}
\author{J.Y. Mutus}
\affiliation{Google, Santa Barbara, CA 93117}
\author{M. Neeley}
\affiliation{Google, Santa Barbara, CA 93117}
\author{C. Neill}
\affiliation{Department of Physics, University of California, Santa Barbara, CA 93106-9530}
\author{P. Roushan}
\affiliation{Google, Santa Barbara, CA 93117}
\author{D. Sank}
\affiliation{Google, Santa Barbara, CA 93117}
\author{A. Vainsencher}
\affiliation{Google, Santa Barbara, CA 93117}
\author{J. Wenner}
\affiliation{Department of Physics, University of California, Santa Barbara, CA 93106-9530}
\author{T.C. White}
\affiliation{Google, Santa Barbara, CA 93117}
\author{John M. Martinis}
\email{martinis@physics.ucsb.edu}
\affiliation{Department of Physics, University of California, Santa Barbara, CA 93106-9530}
\affiliation{Google, Santa Barbara, CA 93117}
\date{\today}

\begin{abstract}

Josephson junctions form the essential non-linearity for almost all superconducting qubits.  The junction is formed when two superconducting electrodes come within $\sim$1 nm of each other.  Although the capacitance of these electrodes is a small fraction of the total qubit capacitance, the nearby electric fields are more concentrated in dielectric surfaces and can contribute substantially to the total dissipation.  We have developed a technique to experimentally investigate the effect of these electrodes on the quality of superconducting devices.  We use $\lambda$/4 coplanar waveguide resonators to emulate lumped qubit capacitors.  We add a variable number of these electrodes to the capacitive end of these resonators and measure how the additional loss scales with number of electrodes. We then reduce this loss with fabrication techniques that limit the amount of lossy dielectrics.  We then apply these techniques to the fabrication of Xmon qubits on a silicon substrate to improve their energy relaxation times by a factor of 5.

\end{abstract}
\maketitle

Josephson junction (JJ) based superconducting qubits are a promising platform for quantum information processing.\cite{kelly2015state, chow2014implementing, poletto2012entanglement, fowler2012surface}  Based on current performances a powerful and error protected processor will require millions of physical qubits.\cite{fowler2012surface}  This number depends strongly on the error rates of individual physical qubits.  The ratio of gate operation time to qubit energy relaxation time ($T_1$) sets a limit on the operation's fidelity.  A dominant source of decoherence in superconducting devices is stray coupling to two level states (TLSs) in amorphous dielectrics.\cite{martinis2005decoherence, quintana2017observation, megrant2012planar, dial2016bulk}  Strong coupling to TLS defects near their resonant frequency also create `holes' in tunable qubit spectra which restrict the frequencies available for control operations.\cite{barends2013coherent}  Nanofabrication techniques such as photolithography, thin film deposition, and etching tend to leave 1-10 nanometer thick interfacial dielectric films with loss tangents on the order of $10^{-3}$, which are difficult to remove.\cite{quintana2014characterization}  Large strides have been made to reduce the energy loss to these surface dielectrics leading to greatly enhanced $T_1$'s in planar and three dimensional superconducting resonators.\cite{megrant2012planar, reagor2013reaching}  While qubit $T_1$'s have also improved, they still lag behind those of resonators and appear to be limited by other loss channels. 

A key difference between resonator and qubit circuits is the inclusion of JJs.  Transmon qubits in particular have two main circuit elements: a linear capacitor and a non-linear inductive element typically constructed from one or more Al/AlO$_x$/Al JJs.  To create capacitors of the highest quality, planar capacitor fabrication can be separated from JJ fabrication and defined identically to planar resonators with critical dimensions of tens of microns.\cite{barends2013coherent, kelly2015state, barends2014logic, corcoles2015demonstration}  Sub-micron JJs are then shadow evaporated and electrically shorted to the capacitor in a lift-off process.  This processing is known to leave behind lossy dielectrics at the surrounding interfaces.\cite{quintana2014characterization}  Furthermore, electrodes that connect to JJs come within close proximity of each other, concentrating electric fields in nearby interfaces and increasing loss.\cite{wang2015surface} Thus, a large amount of loss can come from a relatively small amount of lossy material.  

\begin{figure}[h!]
\begin{centering}
\includegraphics{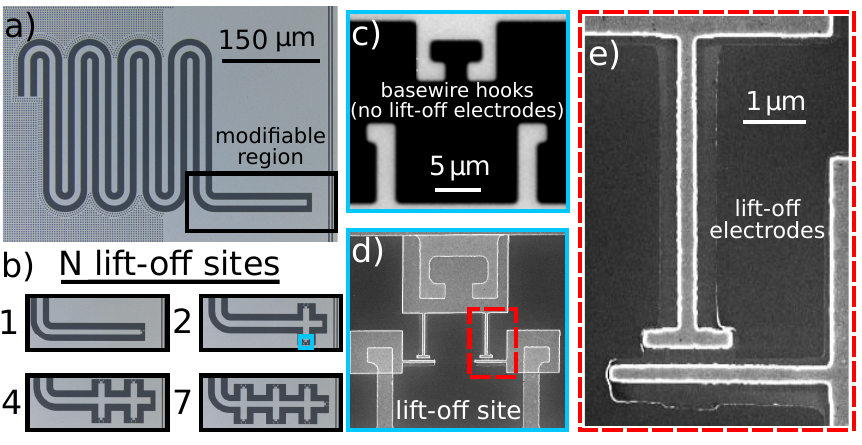} 
\par\end{centering}
\caption{
 (a) Optical micrograph of a `hanging' $\lambda$/4 CPW resonator capacitively coupled to a feedline, highlighting the capacitive end we modify to test the loss from different numbers of JJ electrodes. 
 (b) Optical micrographs showing addition of various numbers of lift-off sites modifying the voltage anti-node of the resonator.
 (c) Zoom in of a single set of basewire `hooks'.  In our standard shadow deposition of JJs we use these hooks to ensure a continuous metal film connects the lift-off JJ electrodes to the basewire metal.  
 (d)  SEM image of a single set of basewire hooks with lift-off JJ electrodes attached.
 (e)  SEM image of electrodes showing damaged nearby substrate. Note that there is no dc contact between the two JJ electrodes.
}

\label{figure:lift-off site} 
\end{figure}

To quantify the additional loss associated with JJ electrodes we mimic the fabrication of JJs connected to high quality coplanar waveguide (CPW) resonators.  We add small capacitive `lift-off sites' to the open end of $\lambda /4$ CPW resonators where there is an electric field anti-node.  These lift-off sites emulate the electrodes leading up to but not including the pair of JJs in the superconducting quantum interference device (SQUID) at the base of Xmon transmon qubits (Fig. \ref{figure:lift-off site}(c/d)).  We measure how loss scales with multiple lift-off sites, effectively amplifying the loss over the background loss of the resonator.  We quantify the loss associated with JJ fabrication and use this knowledge to guide improvements in the fabrication leading directly into improvements in Xmon $T_1$'s.  The most dramatic improvements occur in Xmons fabricated on silicon substrates, where we see an increase in $T_1$ by a factor of 5.

We fabricate these resonators with aluminum base wiring on high resistivity ($>$10 k$\Omega \cdot$cm) intrinsic (100) plane silicon substrates.  Prior to loading for deposition, we sonicate bare wafers in acetone then isopropanol and rinse with deionized (DI) water.  We then dip the wafers in a heated piranha solution, followed by a DI water, then buffered HF to remove the native oxide.  Immediately after blow drying with nitrogen, we load the wafers into a high vacuum electron beam deposition tool and deposit 100 nm of aluminum.  We pattern the coplanar wave guide (CPW) resonators, microwave feedline, and launch pads using optical lithography. We develop the resist in AZ 300 MIF (2\% TMAH in water) developer. We take advantage of the fact that the developer attacks aluminum to wet etch the pattern in the same step as development.

Next we use electron beam lithography (EBL) to pattern electrodes which mimic JJs'.  First, we optically pattern lift-off gold alignment marks, then we pattern and develop the JJ electrode sites using EBL.\cite{barends2013coherent}  We then load the wafer back into the same high vacuum deposition tool used for base aluminum. We use an in-situ 400 V, 0.8 mA/cm$^2$ argon ion mill to etch away the native AlO$_{x}$ to make DC contact between the etched base wire metal and the lift-off metal. \cite{barends2013coherent, quintana2014characterization, chen2014fabrication}  We immediately deposit aluminum at normal incidence leaving a $\sim$ 300 nm gap between ground and signal electrodes which mimic the JJ wiring excluding the JJ itself. It is important to note that we do not make DC electrical contact between the resonator's center trace and the ground plane.  This gap as well as ion mill redeposited residue is displayed using a scanning electron microscope (SEM) image in Fig. \ref{figure:lift-off site}(e).  A more complete description of fabrication procedures is contained within the supplement.\cite{supplement}

\begin{figure}[h!]
\begin{centering}
\includegraphics{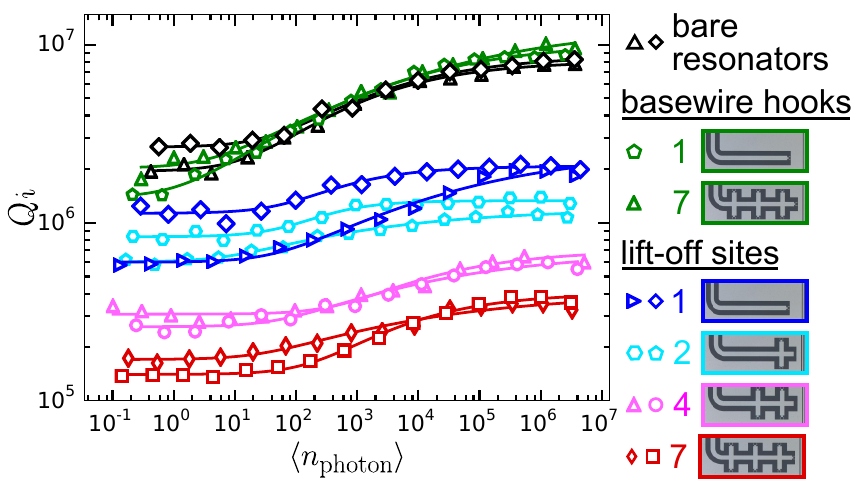} 
\par\end{centering}
\caption{
Representative $Q_i$ measurements versus the average number of photons expected in each modified CPW resonator, at a given drive power.  These measurements are extracted from $S_{21}$ measurements of the shared feedline near the individual resonances.\cite{megrant2012planar}  We fabricate resonators with between zero to seven lift-off sites.  While the low power $Q_i$ of resonators with basewire hooks added (one and seven shown) is relatively unaffected, adding lift-off electrodes lowers $Q_i$ dramatically.
}

\label{figure:resonator} 
\end{figure}

We show an example of a resonator structure in Fig. \ref{figure:lift-off site}(a). This resonator is one of ten CPW $\lambda / 4$ resonators per chip, each capacitively coupled in parallel to a common feed-line.  All resonators are designed with center trace width $w = 24 \ \mu $m, gap to the ground plane ($g$) on either side $g=w$, and resonant frequencies ($f_0$) between 5.5 and 6.0 GHz.  We fabricate CPW resonators with between zero and seven lift-off sites at the voltage anti-node to test the scaling of the additional loss with number of lift-off sites (Fig. \ref{figure:lift-off site}(b)).  The additional lift-off site structures (Fig. \ref{figure:lift-off site}(b-d)) modify the circuit parameters slightly  by adding a parallel capacitance to the open end, but this effect is small ($\sim$1 \% of the total resonator capacitance added per lift-off site). \cite{supplement}  We cool these resonators in a heavily filtered\cite{barends2011minimizing} adiabatic demagnetization refrigerator with a base temperature of 50 mK and extract their internal quality factor, $Q_i = 2 \pi f_0 T_1$ by measuring and fitting the resonators' scattering parameters versus frequency.\cite{megrant2012planar, chiaro2016dielectric}  In Fig. \ref{figure:resonator} we plot this measured $Q_i$ as a function of average photon occupation in the resonator. The low power plateau (around a single average photon occupation) approximates the loss experienced by qubits which also operate at a single photon.

The low power $Q_i$'s of the bare resonators (with no lift-off sites) are within the device-to-device variation of bare resonators fabricated separately (between 2 and 3 million).  These witnesses indicate that the fabrication process itself has little affect on the quality of bare resonators that are buried by resist during the lift-off processes.  Additional base wire `hooks' are used to connect lift-off to base wire aluminum (Fig. \ref{figure:lift-off site}(c)).  Resonators modified to include these hooks (without performing lift-off) have $Q_i$ near the device-to-device variation of the bare resonators as well, indicating little to no added loss from modifying the base wire in this way.  However, when lift-off metal is added, the asymptotic value of $Q_i$ at low photon occupation scales inversely with the number of lift-off sites, and the additional low power loss ($1/Q_i$) per site is 7.9$\times 10^{-7}$.\cite{supplement} This increase of loss with number of JJ electrodes indicates qubits designed with SQUIDs will be twice as sensitive as those designed with single JJs.

\begin{figure}[h!]
\begin{centering}
\includegraphics{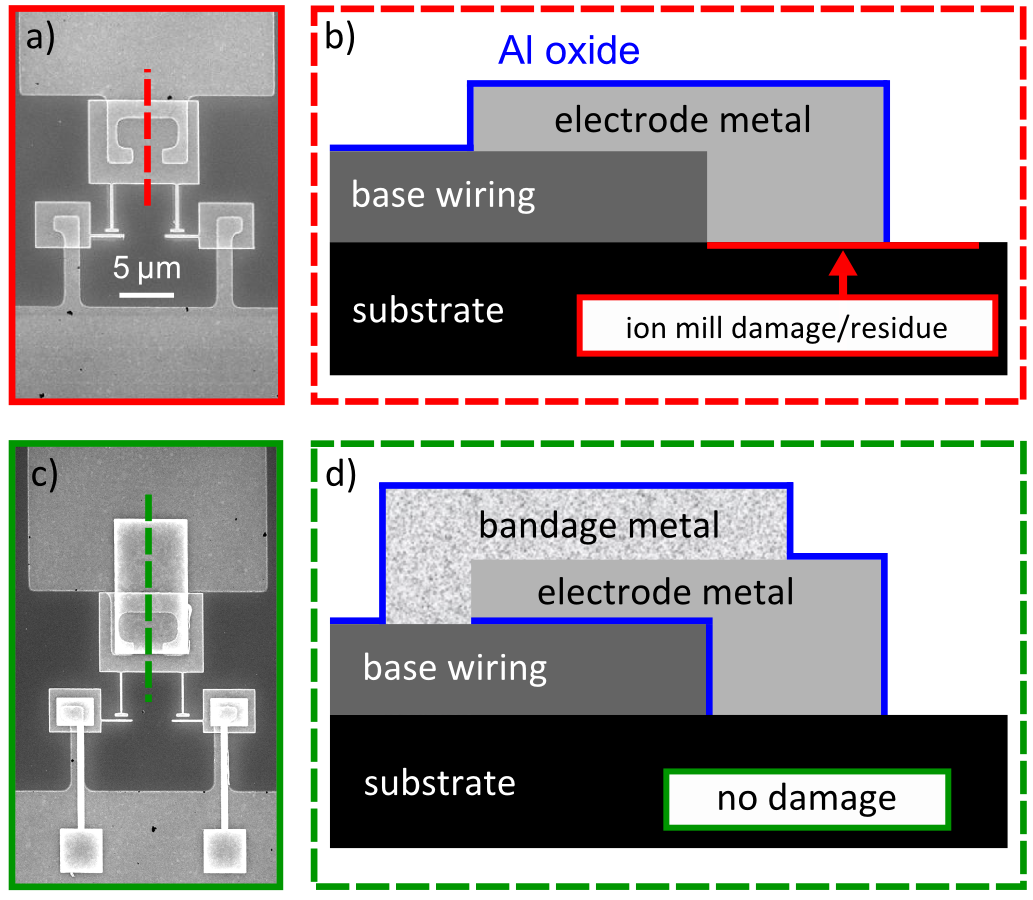} 
\par\end{centering}
\caption{
(a/b) SEM image and cross section cartoon of a lift-off site without bandage metal.  The substrate is damaged during the argon mill process to remove the native oxide from the base wiring. 
(c/d) SEM image and cross section cartoon of lift-off site that has undergone the bandage process, making DC electrical contact between electrode metal and base-wiring while protecting the substrate from the ion mill.}
\label{figure:bandage} 
\end{figure}

The additional loss per lift-off site limits $Q_{i,tot}$ (the total internal quality of the resonator), given by:   

\begin{align} 
\label{eq1}
\frac{1}{Q_{i,tot}} &= \frac{1}{Q_0}+\sum_{j} \frac{p_j}{Q_j},
\end{align}

where $1/Q_0$ is the background loss from other sources, $p_{j}$ (the participation) is the ratio of electric field energy stored in the $j$th volume to the total capacitive energy and $1/Q_{j}$ is the dielectric loss of the $j$th volume.  The volume of dielectric between the electrodes is small due to their close proximity, and this small volume may not always contain TLSs sufficiently near resonance to contribute loss.\cite{martinis2005decoherence} In Fig. \ref{figure:resonator} we see evidence of this effect when only a single pair of JJ electrodes is added the extracted single photon $Q_i$ varies by over a factor of two between resonators at slightly different frequencies. Adding more sites scales the field energy stored near these electrodes and thus loss due to these interfaces approach a `loss tangent regime' where Eq. \ref{eq1} is valid.  With enough lift-off sites the total loss of the resonator is dominated by this added loss and will be less dependent on individual TLS fluctuators.

Previous work has shown that aggressive milling of the substrate leads to amorphization and thus added loss at the substrate-metal (SM) and nearby substrate-vacuum (SV) interfaces on sapphire substrates.\cite{quintana2014characterization}  Using similar experiments on a silicon substrate we found this aggressive ion milled lift-off leaves a roughly 3.9 nm thick interfacial layer underneath the metal.  Using cross-sectional finite-element simulations assuming a relative permittivity $\epsilon_r = 11.6$ for this layer we extract an intrinsic TLS loss tangent $1/Q_{\mathrm{TLS}}=\delta_{\mathrm{TLS}} \sim 7 \times10^{-3}$.\cite{supplement}

\begin{figure}[h!]
\begin{centering}
\includegraphics{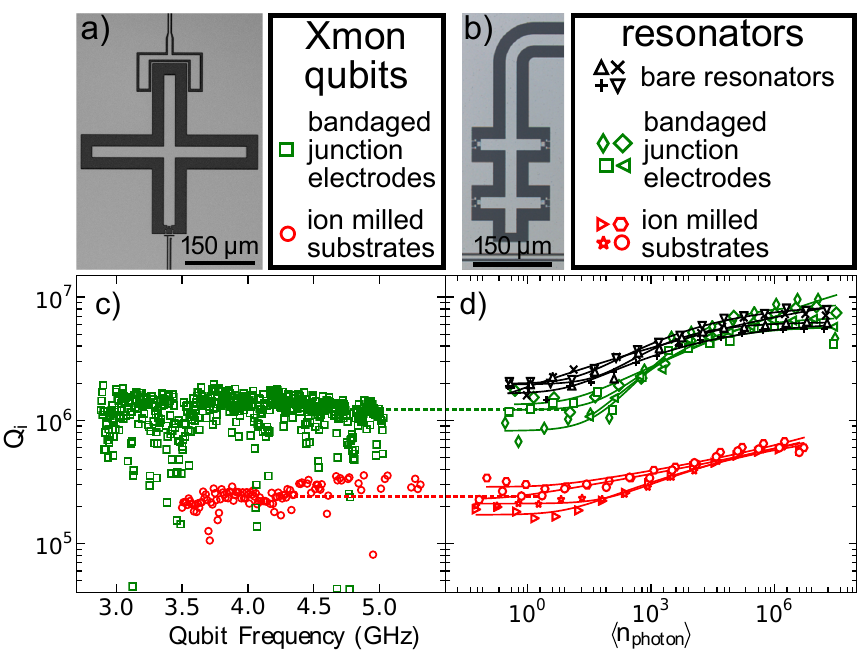} 
\end{centering}
\caption{
(a) Optical micro-graph of typical Xmon qubit capacitively coupled to readout resonator above, and inductively coupled to SQUID bias line below.
(b) Optical micrograph showing capcitive end of a resonator with 4 added lift-off sites.
(c) Effective internal quality factor ($2 \pi f T_1$) of Xmon qubits fabricated on a silicon substrate with and without the bandage process to protect the nearby substrate during ion milling.  Cross capacitor CPW dimensions match the resonators ($w=g=24$ $\mu$m).  Horizontal dashed lines show predicted qubit quality by averaging low power $Q_i$'s of lift-off site resonators prepared with the same fabrication method and nearly equivalent participation.\cite{supplement}
(d) Extracted internal quality factor vs average photon number in the resonator. Notice that when the substrate underneath the lift-off metal is not damaged by the aggressive ion mill, the low power internal quality factor is restored to almost that of the bare resonator. 
}
\label{figure:qubit} 
\end{figure}

To avoid creating this lossy interfacial layer while still making DC contact to the base wiring we break the electrode deposition into two distinct steps: `wiring' and `bandage'.  In the wiring step, we use EBL to define the same JJ electrodes as before, but without any in-situ argon mill.  We use a downstream oxygen asher to descum the developed resist prior to loading the wafer for deposition.  It has been shown that with no in-situ cleaning a de-scum prior to loading can be used to remove any left over contamination from the development process \cite{koppinen2007complete, pop2012fabrication} and reduce loss at interfaces.\cite{quintana2014characterization}  After aluminum deposition we lift-off the resist and unwanted metal in solvents.  Second, we use the bandage step to make galvanic contact between the newly deposited lift-off metal and the base wiring.  We perform a second round of EBL lift-off, but only expose both metal surfaces (base wiring and JJ electrode).  We aggressively ion mill (as detailed above) prior to the bandage deposition.  The substrate is protected by resist and the new bandage metal electrically shorts the JJ electrodes to the base wiring (Fig. \ref{figure:bandage}(c/d)).  The lithography for the bandage metal can also be done optically and this process has been implemented during the JJ fabrication for Xmon, gmon \cite{chen2014qubit}, and fluxmon\cite{quintana2017observation} qubits.

To test the improvement in quality from the bandage process we fabricate resonators with 4 lift-off sites and measure the resulting Q$_{i}$'s (Fig. \ref{figure:qubit}(c/d)).  The low power Q$_{i}$'s are raised back up to around $1.0 \times 10^6$ (a factor of around 5 improvement from the lift-off damaged sites).  A typical Xmon capacitor geometry is shown in Fig. \ref{figure:qubit}(a).  In Fig. \ref{figure:qubit}(b) we plot $Q_i$ corresponding to $T_1$ measurements \cite{supplement} of qubits that underwent both styles of fabrication.  The average $Q_i$ for these qubits is representative of qubits after these processes.  The total capacitances of Xmons and the resonators indicate that resonators with 4 lift-off sites have similar lift-off site participation as qubits \cite{supplement}, and the magnitude of improvement is consistent with the average of low power internal quality factor measurements of these resonators (Fig. \ref{figure:qubit}(c)).

Although the bandage process improves $Q_i$ greatly, there still appears to be residual loss caused by the JJ electrodes limiting $Q_i$ below that of the bare resonators, indicating further improvements are possible.   There are also sections of the bandaged qubit spectrum where the $Q_i$ drops far below it's average value.  These holes occur as the qubits transition frequency ($f_{10}$) is tuned through the resonance of a particularly strongly coupled  TLS.\cite{barends2013coherent}  The tunability of the qubit allows for probing loss as a function of frequency while each resonator only probes at a single frequency. 

In summary, we modified resonators to use them as a tool to directly measure the added capacitve loss from JJ electrodes necessary for most superconducting qubits.  We used this tool to measure different JJ electrode fabrication techniques and found that aggressive ion milling of silicon substrates adds substantial loss. We developed an improved process where we protect the substrate from aggressive ion milling without altering the high coherence capacitor fabrication. We fabricated qubits using this process and measured the average $Q_i$ increase by a factor of 5.  

This work was supported by Google. C. Q. and Z. C. acknowledge support from the National Science Foundation Graduate Research Fellowship under Grant No. DGE-1144085. Devices were made at the UC Santa Barbara Nanofabrication Facility, a part of the NSF funded National Nanotechnology Infrastructure Network.

\bibliographystyle{apsrev}
\bibliography{SiLoRes}
\end{document}


\title{Supplementary Material for "Characterization and Reduction of Capacitive Loss Induced by Sub-Micron Josephson Junction Fabrication in Superconducting Qubits"}

\author{A. Dunsworth}
\thanks{These authors contributed equally to this work}
\affiliation{Department of Physics, University of California, Santa Barbara, CA 93106-9530}
\author{A. Megrant}
\thanks{These authors contributed equally to this work}
\affiliation{Google, Santa Barbara, CA 93117}
\author{C. Quintana}
\affiliation{Department of Physics, University of California, Santa Barbara, CA 93106-9530}
\author{Zijun Chen}
\affiliation{Department of Physics, University of California, Santa Barbara, CA 93106-9530}
\author{R. Barends}
\affiliation{Google, Santa Barbara, CA 93117}
\author{B. Burkett}
\affiliation{Google, Santa Barbara, CA 93117}
\author{B. Foxen}
\affiliation{Department of Physics, University of California, Santa Barbara, CA 93106-9530}
\author{Yu Chen}
\affiliation{Google, Santa Barbara, CA 93117}
\author{B. Chiaro}
\affiliation{Department of Physics, University of California, Santa Barbara, CA 93106-9530}
\author{A. Fowler}
\affiliation{Google, Santa Barbara, CA 93117}
\author{R. Graff}
\affiliation{Google, Santa Barbara, CA 93117}
\author{E. Jeffrey}
\affiliation{Google, Santa Barbara, CA 93117}
\author{J. Kelly}
\affiliation{Google, Santa Barbara, CA 93117}
\author{E. Lucero}
\affiliation{Google, Santa Barbara, CA 93117}
\author{J.Y. Mutus}
\affiliation{Google, Santa Barbara, CA 93117}
\author{M. Neeley}
\affiliation{Google, Santa Barbara, CA 93117}
\author{C. Neill}
\affiliation{Department of Physics, University of California, Santa Barbara, CA 93106-9530}
\author{P. Roushan}
\affiliation{Google, Santa Barbara, CA 93117}
\author{D. Sank}
\affiliation{Google, Santa Barbara, CA 93117}
\author{A. Vainsencher}
\affiliation{Google, Santa Barbara, CA 93117}
\author{J. Wenner}
\affiliation{Department of Physics, University of California, Santa Barbara, CA 93106-9530}
\author{T.C. White}
\affiliation{Google, Santa Barbara, CA 93117}
\author{John M. Martinis}
\email{martinis@physics.ucsb.edu}
\affiliation{Department of Physics, University of California, Santa Barbara, CA 93106-9530}
\affiliation{Google, Santa Barbara, CA 93117}
\date{\today}

\begin{abstract}

We provide supplementary data and calculations.

\end{abstract}

\maketitle

\section{Loss Per Site With and Without Bandage Metal}

\begin{figure}[h!]
\begin{centering}
\includegraphics{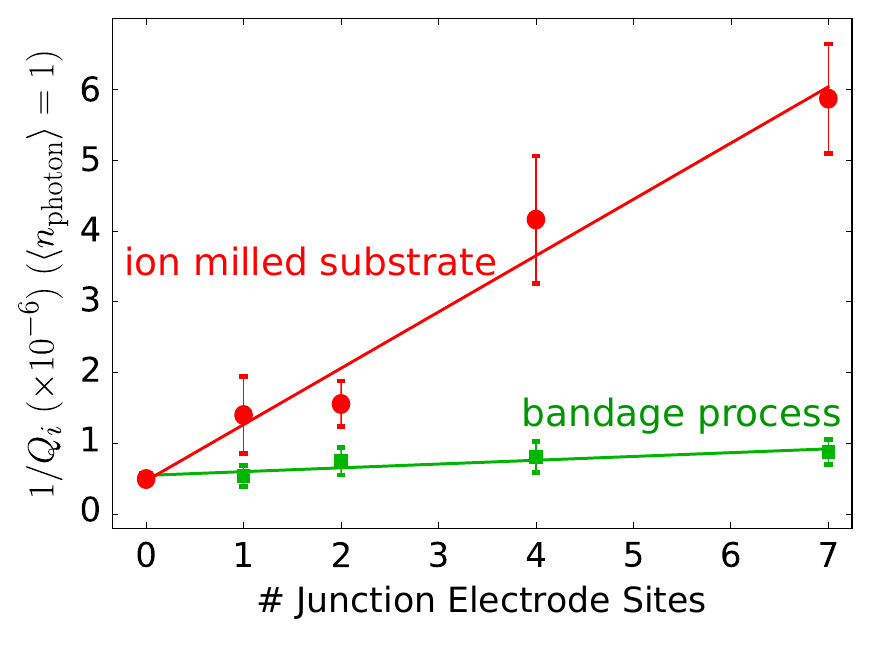}
\par\end{centering}
\caption{Resonator loss ($1/Q_i$) per electrode site is plotted.  Each point and error bar represents the average and standard deviation of at least 4 different resonators on two separate chips. We extract a loss per site (from the line of best fit) for the ion milled substrate lift-off electrodes at a single average photon to be 7.91$\times 10^{-7}$, while the same fit for the bandaged process is 5.35$\times 10^{-8}$.}
\label{figure:I}
\end{figure}

Loss at low power scales linearly with the number of Josephson junction (JJ) electrode sites, as shown in Fig. \ref{figure:I}. We fit this linear relation to extract the loss per JJ electrode site.  We see that the loss is much greater per site for the ion milled substrate process (7.91$\times 10^{-7}$) than for the bandaged process (5.35$\times 10^{-8}$). Lines of best fit effectively average away effect of device to device variation as well as background resonator $Q_i$ to better extract the loss from a single site.  While this experiment stopped at 7 sites, this is an arbitrary upper bound set by design constraints.  More sites could be added to amplify the loss further. Plotted points and error bars represent the average and standard deviation of at least 4 different resonators on two separate chips.

\section{Qubit $T_1$ Data}

Here we plot the measured $T_1$ data for extracted qubit $Q_i$ plotted in the main paper.  We see a stark increase in $T_1$ in qubits processed using the bandage method. The average $T_1$ for the bandage junction qubit is 49.4 $\mu$s whereas the average for the ion milled substrate qubit is 9.5 $\mu$s.

\begin{figure}[h!]
\begin{centering}
\includegraphics{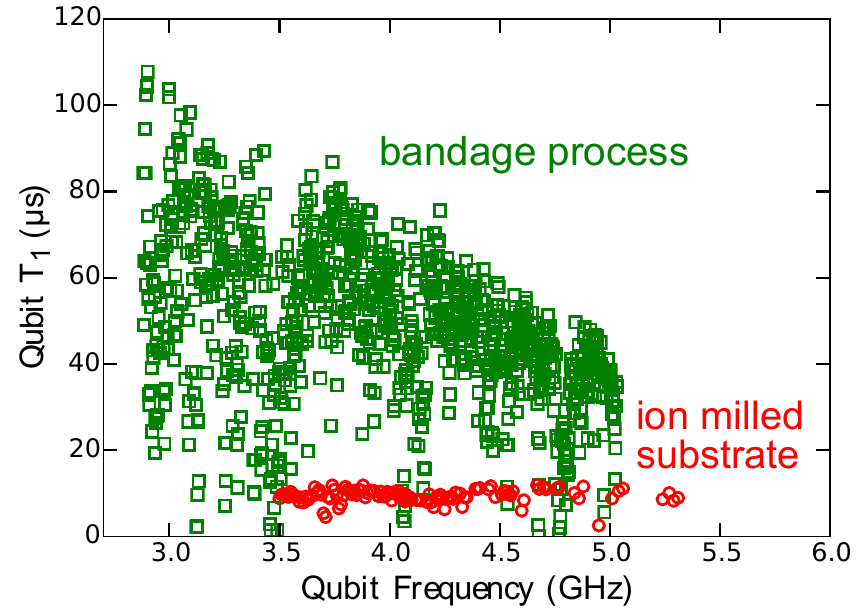} 
\par\end{centering}
\caption{Representative qubit $T_1$ versus frequency for the two differnt junction processing techniques described in the main paper
}
\label{figure:IIII} 
\end{figure}

\section{Lift-off Resonator Center Trace Experiment}

\begin{figure}[h!]
\begin{centering}
\includegraphics{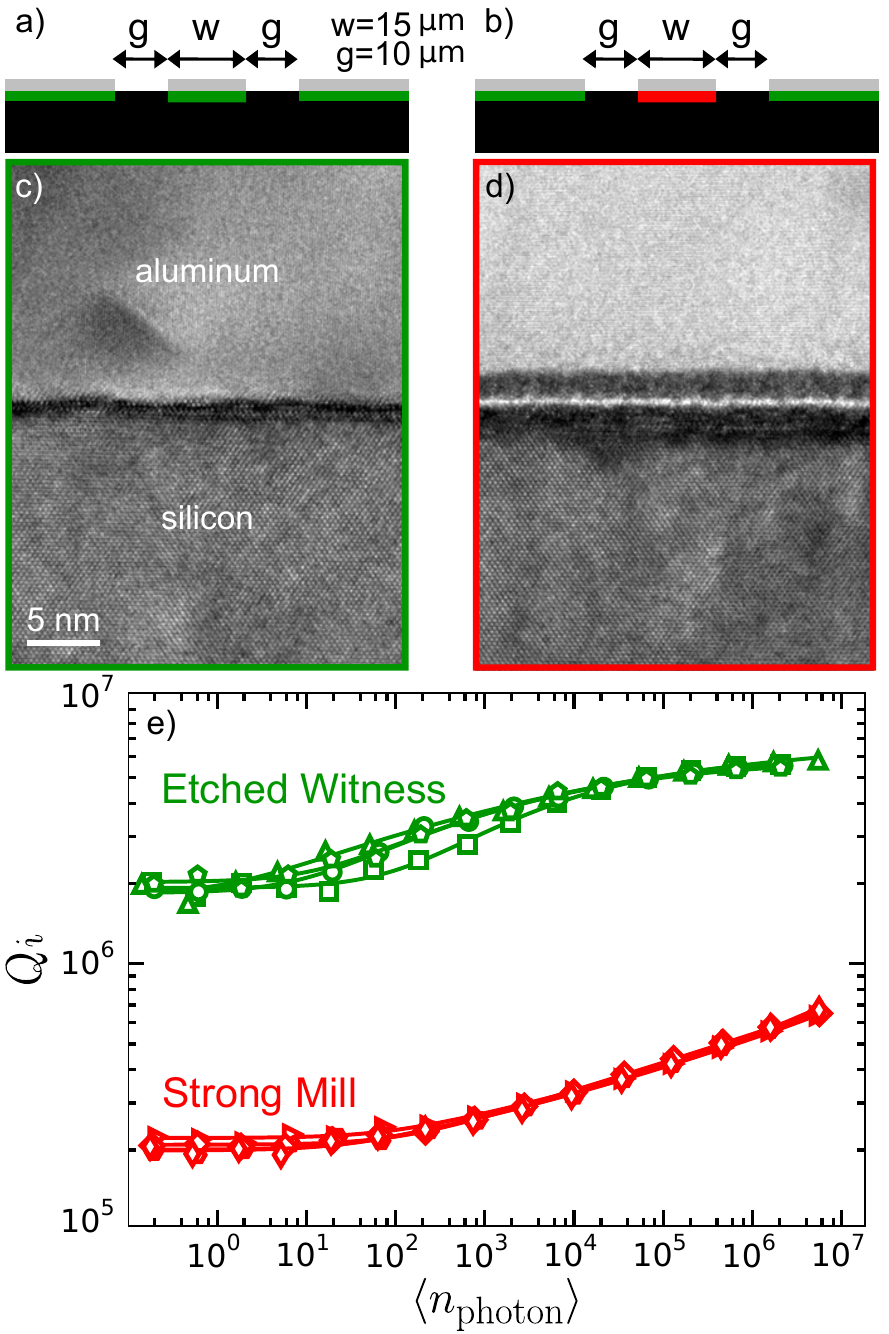} 
\par\end{centering}
\caption{
(a/b) Cartoon cross section of resonators where the center trace is defined with etch/lift-off deposited after strong ion mill. The ground plane metal is deposited on HF dipped silicon and defined by dry etch.
(c/d) Edge on cross-sectional HRTEM images of the substrate-metal interface under center traces where metal was deposited on HF dipped silicon/lifted off after the strong argon ion mill respectively.
(e) $Q_i$ of etched resonator witnesses and resonators where the entire center trace was lifted-off using the junction lift-off process with strong ion mill.}
\label{figure:II} 
\end{figure}

We performed an experiment where the entire center trace of the resonator is defined using lift-off, similar to work done before on a sapphire substrate. \cite{quintana2014characterization}  This experiment isolates the loss from the SM and nearby SV interfaces in a readily simulatable geometry. Basewire preparation was different for these resonators than in the main paper.  The bare wafer was prepared in the same way as described in the main paper, but was then loaded into a Ultra High Vacuum ($P_{base} = 6 \times 10^{-10}$ Torr) molecular beam epitaxy system.  We heated the wafer in vacuum to $\sim 900$ $^{\circ}$C and witnessed $2 \times 1$ reconstruction.  We then allowed the wafer to cool to room temperature in vacuum before depositing 100 nm of Aluminum using an attached electron beam deposition tool.  We use photolithography and a dry etch to define the $\lambda$/4 CPW control resonators as well as $w+2g$ wide holes where lift-off center trace resonators will be formed.  Gold alignment marks were then defined using optical lithography and lift-off.  We define the center traces for the lift-off resonators using EBL lift-off along with the aggressive ion mill treatment detailed in the main paper.

We display cartoon cross-sections of the etch defined and lift-off defined center trace resonators in Fig. \ref{figure:II}(a/b).  We show a cross sectional high-resolution transmission electron microscopy (HRTEM) images of the substrate-metal interface of metal deposited on HF dipped silicon and metal deposited after lift-off processing with the strong ion mill in Fig. \ref{figure:II}(c/d).  We show the internal quality factors of the lift-off center trace resonators and etched witness resonators in Fig. \ref{figure:II}(e).  

\begin{center}
\begin{tabular}{c | c | c | c | c}
 & w  & g  & $Q_i$  & $Q_i$ \\
\textbf{substrate} & ($\mu$m) & ($\mu$m) & (etched) & (ion milled lift \\
~ & ~ & ~ & ~ & -off center trace) \\
\hline
~ & ~ & ~ & ~ & ~ \\
sapphire & 15 & 10 & $6\times 10^5$ & $3\times 10^5$ \\
silicon & 15 & 10 & $2\times 10^6$ & $2\times 10^5$ \\
\end{tabular}
\end{center}

We note a factor of 10 in single photon $Q_i$ between etched and lift-off center trace resonators.  We assume the interfacial layer between HF dipped silicon and aluminum in Fig. \ref{figure:II}(c) is due to the apparent roughness of the silicon surface $\sim$ 1.6 nm.  If we assume this roughness is not changed by ion milling \cite{quintana2014characterization}, we can estimate the interfacial layer underneath the lifted off center trace (Fig. \ref{figure:II}(d)) is 3.9 nm thick. Furthermore, if all excess loss is assumed to come from the SM interface of the center trace, we extract an intrinsic TLS loss tangent $\delta_{\mathrm{TLS}} \sim 7 \times10^{-3}$ for this layer using 2 dimensional finite-element simulations and assuming a relative permittivity $\epsilon_r = 11.6$.

\section{Simulation}

Extensive work has been done to simulate participation of interfaces near junction electrodes.\cite{wang2015surface} We emulate this method to extract expected surface participations for the electrode geometry used in the main paper.  We use a finite element solver (COMSOL) to model the electric field density.\cite{wenner2011surface} Whereas a semi-infinite CPW geometry is efficiently modeled in 2 dimensions, a full 3 dimensional treatment is required to understand how the interfaces participate in an arbitrary geometry. \cite{wang2015surface} We exclude the 1 $\mu $m$^2$ where the actual junction would be be located in simulations.  Individual two level fluctuators need to be considered more carefully in such a small volume as the assumptions that lead to a material with a loss tangent become invalid. \cite{martinis2005decoherence, stoutimore2012josephson}

\begin{figure}[h!]
\begin{centering}
\includegraphics{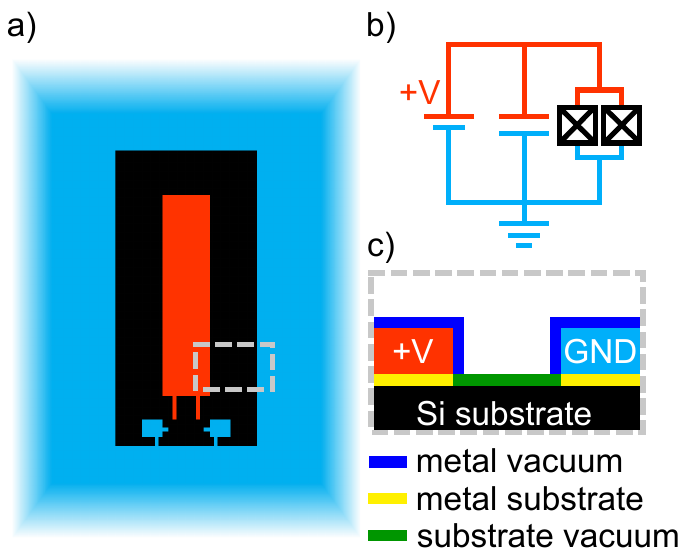}
\par\end{centering}
\caption{
(a)  A cartoon representing the 3 dimensional finite element simulation of an Xmon transmon qubit (the length of the CPW section represents the cross capacitor and is adjusted and extrapolated to match the capacitance in a standard design).  Where a 1 $\mu$m by 1 $\mu$m square around the center of where the junction would be located is removed from both signal and ground.
(b) A circuit representation of the qubit with applied voltage to measure field distributions in simulation.  Junctions are blacked out as they are not simulated.
(c) Cross section indicating interfaces of interest.}
\label{figure:III}
\end{figure}

We apply DC voltage excitation to the center conductor shown in Fig. \ref{figure:III}(a) while the ground plane and corresponding lift-off metal was held at 0 volts.  We display the circuit diagram in \ref{figure:III}(b).  The junctions themselves are displayed in black as they are not considered during the simulation. We show a cross-section indicating the different interfaces in \ref{figure:III}(c).  We calculate the electric field density in the bulk using built-in adaptive meshing. However, to calculate the participations of the three interfaces we apply electro-magnetic boundary conditions as follows:

\begin{align} \label{eq2}
p_{j} &= W^{-1} t_j \epsilon _j \int dA \left| E^2 \right| \\
p_{mv}W/t_{mv} &= \left(1 / \epsilon_{mv} \right) \int dA \left| E_{v,\perp}^2 \right| \\
p_{sm}W/t_{sm} &= \left(\epsilon_s^2 / \epsilon_{sm} \right) \int dA \left| E_{s,\perp}^2 \right| \\
p_{sv}W/t_{sv} &= \epsilon_{sv} \int dA \left| E_{v,\parallel}^2 \right| + \left(1 / \epsilon_{sv} \right) \int dA \left| E_{v,\perp}^2 \right|
\end{align}

All interfaces are assumed to be the same thickness ($t_{mv}=t_{sm}=t_{sv}=3$ nm).  The metal substrate interface is assumed to have $\epsilon_{r,MS}=11.6$ (intrinsic silicon), the substrate vacuum interface is assumed to have $\epsilon_{r,SV}=4.0$ (native silicon oxide) and the metal vacuum interface is assumed to have $\epsilon_{r,MV}=10.0$ (aluminum oxide). The CPW qubit capacitor was assumed to have $w=g=24 \mu$m and the SQUID wires were assumed to be 500 nm wide.  These simulations indicate that around 34\% of the energy stored in surfaces in the entire Xmon qubit is stored in surfaces within $\sim$10 $\mu$m of the junctions.  Combining this knowledge with the increased loss at these interfaces measured in the lift-off resonator center trace experiment, a limit on $Q_{i,tot}$ of the Xmon is set at around $1.4 \times 10^5$, which is similar to what we would predict with lift-off site resonator measurements.

\section{Comparing Xmon Transmon Qubit and Resonator Participations of Junction Lead Sites}

Here we compare how the addition of capacitance through lift-off metal will participate in the trasmon qubit circuit and the resonator circuit. If we have a qubit at 6 GHz with a $w=g=24$ $\mu$m CPW cross capacitor, the length of the capacitor will be roughly:

\begin{align} \label{eq3}
L_{Qc}	&=	160\times4\,\mu\mbox{m} \\
	&=	640\,\mu\mbox{m}
\end{align}

for a 6 GHz CPW $\lambda/4$ resonator on silicon that is also $w=g=24$ $\mu$m the length will be:

\begin{align} \label{eq4}
L_{R}\sim5000\,\mu\mbox{m}
\end{align}

The resonator has a voltage profile extending from the open circuit end to the shorted end of the form:

\begin{align} \label{eq5}
V_{L_{R}}\left(x\right)=V_{0}\cos\left[\frac{\pi x}{2L}\right]
\end{align}

So, the total energy stored capacitively in the resonator is:

\begin{align} \label{eq6}
E_{R}=\frac{1}{2}\int C_{L}V_{R}^{2}dx
\end{align}

Now further we assume that there is a constant geometric capacitance per unit length along the whole length of the resonator and so:

\begin{align} \label{eq7}
E_{R}	&=	\frac{C_{L}V_{R}^{2}}{2}\int_{0}^{L_{R}}\cos^{2}\left[\frac{\pi x}{2L}\right] dx \\
 &=	\frac{C_{L}L_{R}V_{R}^{2}}{4}
\end{align} 

We neglect the small voltage profile along the length of the capacitor of the qubit due to its extended nature.  The total length of the qubit capacitor is only $\sim 3\%$ of the wavelength of a 6 GHz oscillation.  The capacitive energy stored in the qubit (only considering the cross) is:

\begin{align} \label{eq8}
E_{Q}	=	\frac{1}{2}\int C_{L}V_{Q}^{2}dx
	=	\frac{C_{L}L_{Q}V_{Q}^{2}}{2}
\end{align}

and so if the resonator has on average one photon in it and the qubit has a single excitation in it, the ratio of the voltages will be:

\begin{align} \label{eq9}
E_{Q}	&=	E_{R} \\
\implies\left(\frac{V_{R}}{V_{Q}}\right)^{2}	&=	2\left(\frac{L_{Q}}{L_{R}}\right) \\
	&\sim	\frac{1}{4}
\end{align}

In these calculations we assume that the resonator's capacitance is not greatly effected by our added lift-off sites.  We simulate these structures using 2.5 dimensional method of moments software (Sonnet) to justify this assumption.  The added capacitance of the lift-off metal is small compared to that of the CPW structures:

\begin{center}
\begin{tabular}{c | c}
\textbf{geometry} & \textbf{equivalent lumped}\\
~ & \textbf{capacitance (fF)} \\
\hline
6 GHz CPW $\lambda$/4 resonator & 338 \\
\hline
typical Xmon cross capacitor & 86 \\
\hline
typical Josepshon junction & ~ \\
(300 nm $\times$ 300 nm)  & 4 \\
\hline
CPW stub structure & 2.27 \\
\hline
lift-off metal & $\sim$ 0.75 \\
\hline
basewire hooks & <0.05
\end{tabular}
\end{center}

This agrees with SPICE simulations comparing the frequencies of bare resonators and lift-off site resonators.  This calculation implies an Xmon qubit will be four times mores sensitive to loss from these junction electrodes than a resonator at the same frequency.

\section{Standard Fabrication}

Silicon wafers are 450 $\mu$m thick. For base wire and lift-off aluminum depositions we use a Plassys high-vacuum electron beam tool with $P_{base}=2 \times 10^{-8}$ Torr. 

Pre-basewire steps include:
\begin{center}
\begin{tabular}{c | c | c | c}
\textbf{chemical} & \textbf{time} & \textbf{sonicated?} & \textbf{temperature} \\
~ & \textbf{(min)} & ~ & ~ \\
\hline
Acetone & 5 & Yes & room temp. \\
IPA & 5 & Yes & room temp. \\
Nano-strip (piranha) & 10 & No & 70 C \\
Buffered HF & 1 & No & room temp. \\
\end{tabular}
\end{center}

In-situ argon ion mill parameters for etching aluminum oxide are:

\begin{center}
\begin{tabular}{c | c | c | c}
\textbf{beam voltage} & \textbf{beam current} & \textbf{argon flow} & \textbf{time} \\
\textbf{(V)} & \textbf{density at sample} & \textbf{(sccm)} & \textbf{(min)} \\
~ & \textbf{(mA/cm$^2$)} & ~ & ~ \\
\hline
400 &  0.8 & 3.6 & 6 \\
\end{tabular}
\end{center}

Our photo lithogrpahy process is as follows:  
\begin{center}
\begin{tabular}{c | c | c }
\textbf{chemical} & \textbf{spin speed} & \textbf{post-spin bake} \\
~ & \textbf{(rpm)} & \textbf{temp/time} \\
~ & ~ & \textbf{(C/min)} \\
\hline
hexamethyldisilazane & 3000 & - \\
(HMDS) & ~ & ~ \\
\hline
i-line resist & 3000 & 90/1.5 \\
(SPR 955-0.9) & ~ & ~ \\
\end{tabular}
\end{center}

The nominal thickness of the resist is $\sim$0.9 $\mu$m. We use a GCA Auto-Stepper 200 with 0.4 second exposure at $\sim$420 mW/cm$^2$ at the wafer to expose the resist.  We then put the wafer on a 110 C hot plate for 60 seconds to improve resist contrast and development stability.  We develop through exposed features in resist after $\sim$60 seconds in developer, but to wet etch, we develop for an additional $\sim$2-3 minutes.  The wet etch rate of the aluminum increases greatly as the un-oxidized silicon becomes exposed during the etch.  This variable rate leads to aluminum etch back of order 200 nm under the resist and feature rounding. This etch back can be as much as 500 nm on the parts of the wafer that etch through first as the thickness of the metal is non-uniform of $\sim$5-10\% across the wafer due to the deposition process.  This concentration of TMAH has a strong selectivity toward etching aluminum over the silicon substrate (>20:1).  This selectivity is important as lift-off structures need to "step up" over this etched edge and significant under-cuts would break the circuits.  The etch rate of the aluminum is electro-chemically enhanced by the silicon-aluminum interface and can be as much as 10 times faster than the etch rate of aluminum on a sapphire (Al$_2$O$_3$) substrate.      
  
We perform lift-off for gold and the junction electrodes as follows.  Post deposition we submerge the wafer with resist and lift-off metal in beaker of heated ($\sim$ 80 C) N-Methyl-2-pyrrolidone (NMP) based resist stripper for 1 hour.  Once most of the material can be removed with mild agitation we then transplant the wafer into a second heated NMP beaker and sonicate.  We remove the NMP by sonicating the wafer in a beaker of isopropynol and then spin the wafer dry at 1500 RPM.

Our electron beam lithogrpahy process is as follows: 
\begin{center}
\begin{tabular}{c | c | c}
\textbf{chemical} & \textbf{spin speed} & \textbf{post-spin bake} \\
~ & \textbf{(rpm)} & \textbf{temp/time} \\
~ & ~ & \textbf{(C/min)} \\
\hline
PMMA co-polymer & ~ & ~ \\
15\% in Ethyl Lactate & 1500 & 160/10 \\
(MAA) & ~ & ~ \\
\hline
Polymethylmethacrylate & ~ & ~ \\
4\% in Anisole & 2000 & 160/10 \\
(PMMA) & ~ & \\
\end{tabular}
\end{center}

The nominal after-bake thickness of the MAA and PMMA are 500 and 300 nm respectively. We then use a JEOL JBX-6300FS EBL system to expose our resist.  We e-beam write the full stack using a 100 kV, 2 nA beam, with a 1500 $\mu$C/cm$^2$ dose, while the MAA under-layer is exposed using a 350 $\mu$C/cm$^2$ dose for an expected undercut of the PMMA of around 200 nm.  We then develop the pattern in 1:3 Methyl isobutyl ketone (MIBK) to isopropynol for 45 seconds, followed by a rinse in isopropynol for 10 seconds, and blow dry with low pressure (10 psi) nitrogen.

The wafer is heated to a maximum temperature of $\lesssim$150 C for 30 seconds during the descum in the downstream oxygen asher (Gasonics).  It is important to note that the e-beam resist is etched during this process ($\sim$50 nm/min) and thus feature sizes are slightly widened.  We shrink our design widths to compensate. 

\section{Measurement Setup}

We wirebond the resonator chip into a machined Al sample box with copper pcb inserts. We mount this on the 50 mK cold stage of an adiabatic demagnetization refrigerator.  We surround this with a light-tight case which shields from stray thermal radiation from the higher stages which induce extrinsic loss in the superconducting film \cite{barends2011minimizing}. We shield the control lines from stray radiation similarly with home-made in-line IR filters at the cold stage. \cite{barends2011minimizing} We surround the light tight box with a mu-metal shield to reduce stray magnetic fields (predominantly arising from the superconducting coils used to polarize the salt block to cool the cold stage itself) to a manageable level of $\lesssim$10 mG.  We thermalize the control lines at the 4 K stage with 3 in-line 20 dB attenuators, and put the output through a circulator (used as an isolator) to a high electron mobility transitor (HEMT) amplifier both at the 4 K stage.  We attach the output of the amplifier to a 3 dB attenuator also at the 4 K stage to reduce standing waves in the line.  We perform a through measurement on the feedline near resonance of the individual $\lambda / 4$ resonators with a standard microwave vector network analyzer (VNA). For a given power we fit the transmitted wave's amplitude and phase near resonance to an analytic formula for the transmission based on four parameter model for the circuit:

\begin{align} \label{eq9}
\tilde{S}^{-1}_{21}(f)	&=	1 + \frac{Q_i}{Q_c^*}e^{i \phi}\frac{1}{1+i2 Q_i (f-f_0)/f_0}
\end{align} 

$\tilde{S}^{-1}_{21}$ is the inverse transmission normalized so far off resonance transmission is 1.  The four fit parameters are Q$_{c}^{*}$ (the renormalized coupling quality factor of the resonator), $\phi$ (the phase difference between the small impedance mismatches to the left and right of the resonator on the central feedline), $f_0$ (the resonance frequency of the resonator), and Q$_{i}$ (the internal quality factor of the resonator).\cite{megrant2012planar} We then sweep the excitation power of the VNA lower and lower and extract the average photon number inside the resonator based on the excitation power, attenuation, loaded quality factor, and frequency as follows:\cite{barends2009photon}

\begin{align} \label{eq10}
\langle n_{photon} \rangle &= \frac{U}{hf_0} \\
~ &\approx \frac{1}{hf_0} \left( \frac{4 Z_{0}}{\pi} \frac{Q_l^2}{Q_c^*} C_{L} L_{R} P_{drive} \right)
\end{align} 

Where U is the internal energy of the resonator, h is Planck's constant, $f_0$ is the fundamental frequency of the resonator, $Z_0$ is the characteristic impedance of CPW, $Q_l = 1/Q_i + 1/Q_c^*$ is the loaded quality factor of the resonator, $C_{L}$ is the CPW capacitance per unit length of the resonator, $L_{R}$ is the total length of the resonator, and $P_{drive}$ is the power of the traveling wave at the feedline where it couples to the resonator (after all the line attenuations).  The single photon limit approximately captures the physics that describes energy loss in superconducting qubits at the same frequency. \cite{megrant2012planar, chiaro2016dielectric}

\bibliographystyle{apsrev}
\bibliography{supplement}